\documentclass{article}
\usepackage{epsfig,apalike}
\usepackage[english]{babel}
\usepackage[intlimits]{amsmath}
\usepackage{amsthm,amssymb,eucal}
\usepackage{latexsym}
\usepackage{psfrag}
\usepackage{makeidx}

\makeindex
\DeclareFontFamily{U}{msb}{}
\DeclareFontShape{U}{msb}{m}{n}{
<5><6><7><8><9> gen *msbm <10><10.95><12><14.4><17.28><20.74><24.88>msbm10}{}
\DeclareSymbolFont{AMSb}{U}{msb}{m}{n}
\DeclareMathSymbol{\bA}{\mathbin}{AMSb}{'101}
\DeclareMathSymbol{\bB}{\mathbin}{AMSb}{'102}
\DeclareMathSymbol{\bC}{\mathbin}{AMSb}{'103}
\DeclareMathSymbol{\bD}{\mathbin}{AMSb}{'104}
\DeclareMathSymbol{\bE}{\mathbin}{AMSb}{'105}
\DeclareMathSymbol{\bF}{\mathbin}{AMSb}{'106}
\DeclareMathSymbol{\bG}{\mathbin}{AMSb}{'107}
\DeclareMathSymbol{\bH}{\mathbin}{AMSb}{'110}
\DeclareMathSymbol{\bI}{\mathbin}{AMSb}{'111}
\DeclareMathSymbol{\bJ}{\mathbin}{AMSb}{'112}
\DeclareMathSymbol{\bK}{\mathbin}{AMSb}{'113}
\DeclareMathSymbol{\bL}{\mathbin}{AMSb}{'114}
\DeclareMathSymbol{\bM}{\mathbin}{AMSb}{'115}
\DeclareMathSymbol{\bN}{\mathbin}{AMSb}{'116}
\DeclareMathSymbol{\bO}{\mathbin}{AMSb}{'117}
\DeclareMathSymbol{\bP}{\mathbin}{AMSb}{'120}
\DeclareMathSymbol{\bQ}{\mathbin}{AMSb}{'121}
\DeclareMathSymbol{\bR}{\mathbin}{AMSb}{'122}
\DeclareMathSymbol{\bS}{\mathbin}{AMSb}{'123}
\DeclareMathSymbol{\bT}{\mathbin}{AMSb}{'124}
\DeclareMathSymbol{\bU}{\mathbin}{AMSb}{'125}
\DeclareMathSymbol{\bV}{\mathbin}{AMSb}{'126}
\DeclareMathSymbol{\bW}{\mathbin}{AMSb}{'127}
\DeclareMathSymbol{\bX}{\mathbin}{AMSb}{'130}
\DeclareMathSymbol{\bY}{\mathbin}{AMSb}{'121}
\DeclareMathSymbol{\bZ}{\mathbin}{AMSb}{'132}

\def\gex#1{\left\langle #1 \right\rangle}
\def\avg#1{\overline{#1}}

\def\vek#1{\boldsymbol{#1}}
\def\ch#1{\chi\rule[-1ex]{0mm}{2ex}_{#1}}

\def\toi{\rightarrow \infty}

\def\d{\mbox{\rm d}}

\def\T{\mbox{\rm T}}

\def\f{f}
\def\u{u}

\def\r{{\vek{r}}}
\def\M{S}
\DeclareMathOperator{\sgn}{sgn}

\def\cK{{\cal K}}
\def\cR{{\cal R}}
\def\Z{{\vek{Z}}}
\def\f{f}
\def\aphi{\avg{\phi}}

\def\KA{{\bK_0}}
\def\KE{{\bK_\infty}}
\def\muw{\widetilde{\mu}}
\newcommand{\EX}{{\sf EX}}
\newcommand{\DM}{{\sf DM}}
\newcommand{\GF}{{\sf GF}}
\newcommand{\SA}{{\sf SA}}

\def\al{\alpha}

\def\de{\delta}
\def\ka{\kappa}
\def\la{\lambda}

\def\si{\sigma}
\def\La{\Lambda}
\def\Ga{\Gamma}

\def\vep{\varepsilon}

\begin{document}
\begin{center}
{\Large\bf Review on
Scale Dependent Characterization of the Microstructure of Porous Media}\\[24pt]
{R. Hilfer}\\[24pt]
{\em ICA-1, Universit{\"a}t Stuttgart, 70569 Stuttgart, Germany\\
Institut f{\"u}r Physik, Universit{\"a}t Mainz, 55099 Mainz, Germany}
\end{center}

\begin{abstract}
{\small
The paper discusses local porosity theory and its relation
with other geometric characterization methods for porous media 
such as correlation functions and contact distributions.
Special emphasis is placed on the charcterization of geometric
observables through Hadwigers theorem in stochastic geometry.
The four basic Minkowski functionals are introduced into local 
porosity theory, and for the first time a relationship is established 
between the Euler characteristic and the local percolation
probabilities.
Local porosity distributions and local percolation probabilities
provide a scale dependent characterization of the microstructure
of porous media that can be used in an effective medium approach
to predict transport.}
\end{abstract}
\vspace*{2cm}

\section{Introduction}

A crucial prerequisite for the prediction of transport
parameters of porous media is a suitable characterization
of the microstructure \cite{dul92,adl92,sah95,hil95d}.
Despite a long history of scientific study
the microstructure of porous media 
continues to be investigated in many areas 
of fundamental and applied research ranging from geophysics
\cite{HN85}, hydrology \cite{mar86,BV87}, petrophysics \cite{lak89}
and civil engineering \cite{ehl95,DE96b} to the materials science
of composites \cite{cri98}.

My primary objective in this article is to review briefly the
application of local porosity theory, 
introduced in \cite{hil91d,hil92a,hil95d},
as a method that provides a scale dependent
geometric characterization of porous or 
heterogeneous media.
A functional theorem of Hadwiger 
\cite[p.39]{had55} emphasizes the importance of four 
set-theoretic functionals for the geometric characterization
of porous media.
In contrast herewith local porosity theory has emphasized
geometric observables, that are not covered by Hadwigers theorem
\cite{hil93b,hil94b,hil94g}.
Other theories have stressed the importance of correlation
functions \cite{TS82,ste85} or contact distributions
\cite{LT92a,MS92,SKM95} for characterization purposes.
Recently advances in computer and imaging technology
have made threedimensional microtomographic images 
more readily available.
Exact microscopic solutions are thereby becoming
possible and have recently been calculated
\cite{hil98h,hil99a,hil99c}.
Moreover, the availability of threedimensional 
microstructures allows to test approximate theories
and geometric models and to distinguish them quantitatively.

Distinguishing porous microstructures in a quantitative fashion 
is important for reliable predictions and it requires 
apt geometric observables.
Examples of important geometric observables are
porosity and specific internal surface area \cite{bea72,dul92}.
It is clear however, that porosity and specific
internal surface area alone are not sufficient
to distinguish the infinite variety of porous 
microstructures.

Geometrical models for porous media may be roughly
subdivided into the classical capillary tube and slit
models \cite{dul92}, grain models \cite{SKM95}, 
network models \cite{fat56a,BMC93}, 
percolation models \cite{CD77,sah93},
fractal models \cite{KT86,RT97},
stochastic reconstruction models \cite{qui84,adl92}
and diagenetic models \cite{RS85,BO97}.
Little attention is usually paid to match
the geometric characteristics of a model
geometry to those of the experimental sample, 
as witnessed by the undiminished popularity
of capillary tube models.
Usually the matching of geometric observables
is limited to the porosity alone.
Recently the idea of stochastic reconstruction models 
has found renewed interest
\cite{adl92,rob97,YT98a}.
In stochastic reconstruction models one tries to match 
not only the porosity but also other geometric quantities 
such as specific internal surface, correlation functions, 
or linear and spherical contact distributions.
As the number of matched quantities increases one
expects that also the model approximates better
the given sample.
Matched models for sedimentary rocks have recently 
been subjected to a quantitative comparison with the 
experimentally obtained microstructures \cite{hil99c}.

\section{Geometrical Problems in Porous Media}

A two-component porous sample $\bS=\bP\cup\bM$ is 
defined as the union of two closed subsets $\bP\subset\bR^3$ 
and $\bM\subset\bR^3$ where $\bP$ denotes the pore space 
(or component 1 in a heterogeneous medium) and $\bM$ 
denotes the matrix space (or component 2).
For simplicity only two-component media will be considered
throughout this paper, but most concepts can be generalized to
media with an arbitrary finite number of components.
A particular pore space configuration may be described
using the characteristic (or indicator) function 
$\ch{\bP}(\r)$ of a set $\bP$.
It is defined for arbitrary sets $\bP$ as
\index{characteristic function}\index{indicator function}
\begin{equation}
\ch{\bP}(\r)=\begin{cases}
                1 &\text{for~}\r\in\bP\\
                0 &\text{for~}\r\notin\bP .
             \end{cases}
\label{charfunc}
\end{equation}
The geometrical problems in porous media arise because 
in practice the pore space configuration $\ch{\bP}(\r)$
is usually not known in detail.
On the other hand the solution of a physical boundary 
value problem would require detailed 
knowledge of the internal boundary, and hence of 
$\ch{\bP}(\r)$.

While it is becoming feasible to digitize samples
of several mm$^3$ with a resolution of a few $\mu$m
this is not possible for larger samples.
For this reason the true pore space $\bP$ is
often replaced by a geometric model $\widetilde\bP$.
One then solves the problem for the model geometry and hopes
that its solution $\widetilde{\u}$ obeys 
$\widetilde{\u}\approx\u$ in some sense.
Such an approach requires
quantitative methods for the comparison
of $\bP$ and the model $\widetilde\bP$.
This in turn raises the problem of finding generally applicable
quantitative geometric characterization methods
that allow to evaluate the accuracy of geometric
models for porous microstructues.
The problem of quantitative geometric characterization
arises also when one asks which geometrical characteristics 
of the microsctructure $\bP$ have the greatest influence on 
the properties of the solution $\u$ of a given boundary value 
problem.

Some authors introduce more than one geometrical model for
one and the same microstructure when calculating different physical 
properties (e.g. diffusion and conduction).
It should be clear that such models make it difficult to extract
reliable physical or geometrical information.

\section{Geometric Characterizations}
\subsection{General Considerations}
\label{generalconsid}
A general geometric characterization of stochastic media
should provide macroscopic geometric observables that
allow to distinguish media with different microstructures
quantitatively.
In general, a stochastic medium is defined
as a probability distribution on a space of geometries
or configurations.
Probability distributions and expectation values 
of geometric observables are candidates for
a general geometric characterization.

A general geometric characterization should fulfill
four criteria to be useful in applications.
\index{gometric characterization}
These four criteria were advanced in \cite{hil95d}.
First, it must be well defined. 
This obvious requirement is sometimes violated.
The so called
``pore size distributions'' measured in mercury
porosimetry are not geometrical observables in the
sense that they cannot be determined from knowledge of 
the geometry alone.
Instead they are capillary pressure curves whose
calculation involves physical quantities such
as surface tension, viscosity or flooding history \
\cite{hil95d}.
\index{pore size distribution}
Second, the geometric characterization should be 
directly accessible in experiments.
The experiments should be independent of the quantities 
to be predicted.
Thirdly, the numerical implementation should not require 
excessive amounts of data.
This means that the amount of data should be manageable 
by contemporary data processing technology.
Finally, a useful geometric characterization should
be helpful in the exact or approximate theoretical
calculations.

\subsection{Geometric Observables}
Well defined geometric observables are the basis for
the geometric characterization of porous media.
A perennial problem in all applications is to identify
those macroscopic geometric observables that are 
relevant for distinguishing between classes of microstructures.
One is interested in those properties of 
the microstructure that influence the macroscopic physical
behaviour.
In general this depends on the details of the physical
problem, but some general properties of the microstructure
such as volume fraction or porosity are known to
be relevant in many situations.
Hadwigers theorem \cite{had55} is an example of
a mathematical result that helps to identify an
important class of such general geometric properties of porous media.
It will be seen later, however, that there exist
important geometric properties that are not
members of this class.

A geometric observable $\f$ is a mapping (functional) that 
assigns to each admissible pore space $\bP$ a real number 
$\f(\bP)=\f(\bP\cap\bS)$
that can be calculated from $\bP$ without solving a physical
boundary value problem.
\index{geometric observable}
A functional whose evaluation requires the solution of a physical
boundary value problem will be called a physical observable.
\index{physical observable}

Before discussing examples for geometric observables 
it is necessary to specify the admissible geometries $\bP$.
The set $\cR$ of admissible $\bP$ is defined as 
the set of all finite unions of compact convex sets
\cite{had55,SKM95,SW92,sch93}.
Because $\cR$ is closed under unions and 
intersections it is called the convex ring.
The choice of $\cR$ is convenient for applications
because digitized porous media can be considered as 
elements from $\cR$
and because continuous observables defined for 
convex compact sets can be continued to all of $\cR$.
The set of all compact and convex subsets of $\bR^d$ 
is denoted as $\cK$.
For subsequent discussions the 
Minkowski addition of two sets $\bA,\bB\subset\bR^d$ 
is defined as
\index{Minkowski summation}
\begin{equation}
\bA+\bB=\{\vek{x}+\vek{y}:\vek{x}\in\bA,\vek{y}\in\bB\}.
\end{equation}
Multiplication of $\bA$ with a scalar is defined by
$a\bA=\{a\vek{x}:x\in\bA\}$ for $a\in\bR$.

Examples of geometric observables are the volume of $\bP$
or the surface area of the internal
$\partial\bP=\partial\bM=\bP\cap\bM$.
\footnote{The boundary $\partial \bG$ of a set $\bG$ 
is defined as the
\index{boundary of a set}
difference between the closure and the interior of $\bG$
where the closure is the intersection of all closed
sets containing $\bG$ and the interior is the union of
all open sets contained in $\bG$.}
Let 
\begin{equation}
V_d(\bK) = \int_{\bR^d}\ch{\bP}(\r)\d^d\r
\end{equation}
denote the $d$-dimensional Lebesgue volume
of the compact convex set $\bK$.
The volume is hence a functional $V_d:\cK\to\bR$ on $\cK$.
An example of a compact convex set is the unit ball 
$\bB^d=\{\vek{x}\in\bR^d:|\vek{x}|\leq1\}=\bB^d(\vek{0},1)$
centered at the origin $\vek{0}$ whose volume is
\begin{equation}
\ka_d=V_d(\bB^d)=\frac{\pi^{d/2}}{\Ga(1+(d/2))} .
\label{volball}
\end{equation}
Other functionals on $\cK$ can be constructed from the volume
by virtue of the following fact.
For every compact convex $\bK\in\cK$ and every
$\vep\geq 0$ there are numbers $V_j(\bK),j=0,\ldots,d$ 
depending only on $\bK$ such that 
\index{Steiners formula}
\begin{equation}
V_d(\bK+\vep\bB^d)=\sum_{j=0}^d V_j(\bK)\vep^{d-j}\ka_{d-j}
\label{steiner}
\end{equation}
is a polynomial in $\vep$.
This result is known as Steiners formula \cite{had55,SKM95}.
The numbers $V_j(\bK),j=0\ldots,d$ define functionals on 
$\cK$ similar to the volume $V_d(\bK)$.
The quantities\
\index{quermassintegral}
\begin{equation}
W_i(\bK)=\frac{\ka_i V_{d-i}(\bK)}{\binom{d}{i}}
\label{querm}
\end{equation}
are called quermassintegrals \cite{sch93}.
From \eqref{steiner} one sees that
\begin{equation}
\lim_{\vep\to 0}\frac{1}{\vep}(V_d(\bK+\vep\bB^d)-V_d(\bK))
=\ka_1 V_{d-1}(\bK),
\label{area}
\end{equation}
and from \eqref{volball} that $\ka_1=2$.
Hence $V_{d-1}(\bK)$ may be viewed as half the surface area.
The functional $V_1(\bK)$ is related to the mean width $w(\bK)$
defined as the mean value of the distance between a pair of
parallel support planes of $\bK$.
\index{mean width $w$}
The relation is 
\begin{equation}
V_1(\bK)=\frac{d\ka_d}{2\ka_{d-1}}w(\bK)
\end{equation}
which reduces to $V_1(\bK)=w(\bK)/2$ for $d=3$.
Finally the functional $V_0(\bK)$ is evaluated from
\eqref{steiner} by dividing with $\vep^d$ and taking
the limit $\vep\toi$.
It follows that $V_0(\bK)=1$ for all $\bK\in\cK\setminus\{\emptyset\}$.
One extends $V_0$ to all of $\cK$ by defining 
$V_0(\emptyset)=0$.
The geometric observable $V_0$ is called Euler characteristic.
\index{Euler characteristic}

The geometric observables $V_i$ have several important properties.
They are Euclidean invariant (i.e. invariant under rigid motions),
additive and monotone.
Let $T_d\cong(\bR^d,+)$ denote the group of translations with
vector addition as group operation and let $SO(d)$ be
the matrix group of rotations in $d$ dimensions \cite{BR86}.
The semidirect product $E_d=T_d\odot SO(d)$ is the Euclidean
group of rigid motions in $\bR^d$. 
\index{Euclidean group}\index{group of rigid motions}
It is defined as the set of pairs $(\vek{a},A)$ 
with $\vek{a}\in\T_d$ and $A\in SO(d)$ and group operation
\begin{equation}
(\vek{a},A)\circ (\vek{b},B)= (\vek{a}+A\vek{b},AB).
\end{equation}
An observable $\f:\cK\to\bR$ is called euclidean invariant
or invariant under rigid motions
\index{euclidean invariant}\index{invariant under rigid motions}
if 
\begin{equation}
\f(\vek{a}+A\bK)=\f(\bK)
\end{equation}
holds for all $(\vek{a},A)\in E_d$ and all $\bK\in\cK$.
Here $A\bK=\{A\vek{x}:\vek{x}\in\bK\}$ denotes the rotation
of $\bK$ and $\vek{a}+\bK=\{\vek{a}\}+\bK$ its translation.
A geometric observable $\f$ is called additive
\index{additive functional}
if
\begin{align}
\f(\emptyset)&=0\\
\f(\bK_1\cup\bK_2)+\f(\bK_1\cap\bK_2)&=\f(\bK_1)+\f(\bK_2)
\end{align}
holds for all $\bK_1,\bK_2\in\cK$ with $\bK_1\cup\bK_2\in\cK$.
Finally a functional is called monotone if for $\bK_1,\bK_2\in\cK$
with $\bK_1\subset\bK_2$ follows $\f(\bK_1)\leq\f(\bK_2)$.
\index{monotone functional}

The special importance of the functionals $V_i(\bK)$ arises from 
the following theorem of Hadwiger \cite{had55}.
A functional $\f:\cK\to\bR$ is euclidean invariant,
additive and monotone if and only if it is a linear
combination
\begin{equation}
\f=\sum_{i=0}^d c_iV_i
\label{Hadwiger}
\end{equation}
with nonnegative constants $c_0,\ldots,c_d$.
The condition of monotonicity can be replaced with
continuity at the expense of allowing also negative $c_i$,
and the theorem remains valid \cite{had55}.
If $\f$ is continuous on $\cK$, additive and euclidean invariant
it can be additively extended to the convex ring $\cR$
\cite{SW92}.
The additive extension is unique and given by the
inclusion-exclusion formula
\begin{equation}
\f\left(\bigcup_{i=1}^m\bK_1\right)=
\sum_{\bI\in{\cal P}(m)}(-1)^{|\bI|-1}
\f\left(\bigcap_{i\in\bI}\bK_i\right)
\label{inclexcl}
\end{equation}
where ${\cal P}(m)$ denotes the family of nonempty subsets
of $\{1,\ldots,m\}$ and $|\bI|$ is the number of elements 
of $\bI\in{\cal P}(m)$.
In particular, the functionals $V_i$ have a unique additive extension
to the convex ring $\cR$\cite{SW92}, which is again be denoted by $V_i$.

For a threedimensional porous sample with $\bP\in\cR$ the extended 
functionals $V_i$ lead to two frequently used geometric observables.
The first is the porosity of a porous sample $\bS$ defined as
\index{porosity}
\begin{equation}
\phi(\bP\cap\bS)=\phi_3(\bP\cap\bS)=\frac{V_3(\bP\cap\bS)}{V_3(\bS)},
\label{porosity}
\end{equation}
and the second its specific internal surface area
which may be defined in view of \eqref{area} as
\index{specific internal surface area}
\begin{equation}
\phi_2(\bP\cap\bS)=\frac{2V_2(\bP\cap\bS)}{V_3(\bS)}.
\label{specsurf}
\end{equation}
The two remaining observables 
$\phi_1(\bP\cap\bS)=V_1(\bP\cap\bS)/V_3(\bS)$ and
$\phi_0(\bP\cap\bS)=V_0(\bP\cap\bS)/V_3(\bS)$ have received less attention
in the porous media literature.
The Euler characteristic $V_0$ on $\cR$ coincides with
the identically named topological invariant.
For \mbox{$d=2$} and $\bG\in\cR$ one has
$V_0(\bG)=c(\bG)-c'(\bG)$ where $c(\bG)$ is the number
of connectedness components of $\bG$, and $c'(\bG)$ denotes
the number of holes (i.e. bounded connectedness components
of the complement).

\subsection{Definition of Stochastic Porous Media}
For theoretical purposes the pore space $\bP$ is frequently
viewed as a random set \cite{SKM95,hil95d}.
In practical applications the pore space is usually discretized
because of measurement limitations and finite resolution.
For the purpose of discussion the set $\bS\subset\bR^3$ is a 
rectangular parallelepiped whose sidelengths are $M_1,M_2$ 
and $M_3$ in units of the lattice constant $a$ (resolution) 
of a simple cubic lattice.
The position vectors 
$\r_i=\r_{i_1\ldots i_d}=(a i_1,\ldots,a i_d)$
with integers $1\leq i_j \leq M_j$ are used to
label the lattice points, and $\r_i$ 
is a shorthand notation for $\r_{i_1\ldots i_d}$.
Let $\bV_i$ denote a cubic volume element (voxel)
centered at the lattice site $\r_i$.
Then the discretized sample may be represented
as $\bS=\bigcup_{i=1}^N\bV_i$.
The discretized pore space $\widetilde{\bP}$, defined as
\begin{equation}
\widetilde{\bP}=\bigcup_{\{i:\ch{\bP}(\r_i)=1\}}\bV_i ,
\end{equation}
is an approximation to the true pore space $\bP$.
For simplicity it will be assumed that the
discretization does not introduce errors,
i.e. that $\widetilde{\bP}=\bP$, and that
each voxel is either fully pore or fully matrix.
This assumption may be relaxed to allow 
voxel attributes such as internal surface or
other quermassintegral densities.
The discretization into voxels reflects the limitations arising
from the experimental resolution of the porous structure.
A discretized pore space for a bounded sample belongs 
to the convex ring $\cR$ if the voxels are convex and 
compact.
Hence, for a simple cubic discretization the pore
space belongs to the convex ring.
A configuration (or microstructure) $\Z$ of a
\index{configuration of porous medium $\Z$}
$2$-component medium may be represented 
in the simplest case by a sequence
\begin{equation}
\Z=(Z_1,\ldots,Z_N)=(\ch{\bP}(\r_1),\ldots,\ch{\bP}(\r_N))
\label{corresp}
\end{equation}
where $\r_i$ runs through the lattice points and $N=M_1M_2M_3$.
This representation corresponds to the simplest discretization
in which there are only two states for each voxel indicating
whether it belongs to pore space or not.
In general a voxel could be characterized
by more states reflecting the microsctructure within
the region $\bV_i$.
In the simplest case there is a one-to-one correspondence 
between $\bP$ and $\Z$ given by \eqref{corresp}.
Geometric observables $\f(\bP)$ then correspond to functions
$\f(\Z)=\f(z_1,\ldots,z_N)$.

As a convenient theoretical idealization it is frequently
assumed that porous media are random realizations
drawn from an underlying statistical ensemble.
A discretized stochastic porous medium is
defined through the discrete probability density
\index{stochastic porous medium}
\begin{equation}
p(z_1,\ldots,z_N) = \mbox{Prob}\{(Z_1=z_1)\wedge\ldots\wedge(Z_N=z_N)\}
\label{stochmed}
\end{equation}
where $z_i\in\{0,1\}$ in the simplest case. 
It should be emphasized that the probability density $p$
is mainly of theoretical interest.
In practice $p$ is usually not known.
An infinitely extended medium or microstructure 
is called stationary or
\index{homogeneous microstructure}
statistically homogeneous if $p$ is invariant under
spatial translations.
\index{stationary porous medium}
It is called isotropic if $p$ is invariant under rotations.
\index{isotropic porous medium}

\subsection{Moment Functions and Correlation Functions}

A stochastic medium was defined through
its probability distribution $p$.
In practice $p$ will be even less accessible than the 
microstructure $\bP=\Z$ itself.
Partial information about $p$ can be obtained 
by measuring or calculating expectation values
of a geometric observable $\f$. These are defined as
\index{expectation values $\langle\ldots\rangle$}
\begin{equation}
\langle f(z_1,\ldots,z_N) \rangle = \sum_{z_1=0}^1\ldots\sum_{z_N=0}^1
f(z_1,\ldots,z_N)p(z_1,\ldots,z_N)
\label{expectation}
\end{equation}
where the summations indicate a summation over all configurations.
Consider for example the porosity $\phi(\bS)$ defined in \eqref{porosity}.
For a stochastic medium $\phi(\bS)$ becomes a random variable.
Its expectation is 
\begin{align}
\gex{\phi}&= 
\frac{\gex{V_3(\bP)}}{V_3(\bS)}
=\frac{1}{V_3(\bS)}\int_{\bS}\gex{\ch{\bP}(\r)}\d^3\r\nonumber\\
&=\frac{1}{V_3(\bS)}\sum_{i=1}^N\gex{z_i}V_3(\bV_i)
=\frac{1}{N}\sum_{i=1}^N\gex{z_i}\nonumber\\
&=\frac{1}{N}\sum_{i=1}^N\mbox{Prob}\{z_i=1\}
=\frac{1}{N}\sum_{i=1}^N\mbox{Prob}\{\r_i\in\bP\} .
\label{avgporosity}
\end{align}
If the medium is statistically homogeneous then
\begin{equation}
\gex{\phi}=\mbox{Prob}\{z_i=1\}=
\mbox{Prob}\{\r_i\in\bP\} = \gex{\ch{\bP}(\r_i)}
\end{equation}
independent of $i$.
It happens frequently that one is given only a single
sample, not an ensemble of samples.
It is then necessary to invoke an ergodic hypothesis
that allows to equate spatial averages with ensemble 
averages.

The porosity is the first member in a hierarchy of 
moment functions.
The $n$-th order moment function is defined generally as
\index{moment functions $\M_n$}
\begin{equation}
\M_n(\r_1,\ldots,\r_n)=\gex{\ch{\bP}(\r_1)\ldots\ch{\bP}(\r_n)}
\end{equation}
for $n\leq N$.
\footnote{
If a voxel has other attributes besides 
being pore or matrix one may define also mixed moment functions
$\M_{i_1\ldots i_n}(\r_1,\ldots,\r_n)=\gex{\phi_{i_1}(\r_1)\ldots\phi_{i_n}(\r_n)}$ where $\phi_i(\r_j)=V_i(\bP\cap\bV_j)/V_i(\bV_j)$ for $i=1,\ldots d$
are the quermassintegral densities for the voxel at site $\r_j$.
}
For stationary media 
$\M_n(\r_1,\ldots\r_n)=g(\r_1-\r_n,\ldots,\r_{n-1}-\r_n)$
where the function $g$ depends only on $n-1$ variables.
Another frequently used expectation value is the correlation function
which is related to $\M_2$.
For a homogeneous medium it is defined as 
\index{correlation function}
\begin{equation}
  \label{eq:g_gen}
  G(\r_0,\r) = G(\r-\r_0) = 
  \frac{\gex{\ch{\bP}(\r_0)\ch{\bP}(\r)}-\gex{\phi}^2}
  {\gex{\phi}(1-\gex{\phi})} =
  \frac{\M_2(\r-\r_0)-(\M_1(\r_0))^2}
  {\M_1(\r_0)(1-\M_1(\r_0))}
\end{equation}
where $\r_0$ is an arbitrary reference point, and $\gex{\phi}=\M_1(\r_0)$.
If the medium is isotropic then $G(\r)=G(|\r|)=G(r)$.
Note that $G$ is normalized such that $G(0)=1$ and $G(\infty)=0$.

The hierarchy of moment functions $\M_n$, similar to $p$, 
is mainly of theoretical interest.
For a homogeneous medium $\M_n$ is a function of $n-1$
variables.
To specify $\M_n$ numerically becomes impractical
as $n$ increases. 
If only $100$ points are required along each coordinate axis
then giving $\M_n$ would require $10^{2d(n-1)}$ numbers.
For $d=3$ this implies that already at $n=3$ it becomes economical
to specify the microstructure $\bP$ directly rather than 
incompletely through moment or correlation functions.

\subsection{Contact Distributions}
An interesting geometric characteristic introduced and
discussed in the field of stochastic geometry are 
contact distributions \cite[p. 206]{del72,SKM95}.
Certain special cases of contact distributions have appeared
also in the porous media literature \cite{dul92}.
Let $\bG$ be a compact test set containing the origin $\vek{0}$.
Then the contact distribution is defined as the conditional
probability
\index{contact distribution}
\begin{equation}
H_\bG(r)=1-\mbox{\rm Prob}\{\vek{0}\notin\bM+(-r\bG)|\vek{0}\notin\bM\}
=1-\frac{\mbox{\rm Prob}\{\bM\cap r\bG=\emptyset\}}{\phi}
\end{equation}
If one defines the random variable $R=\inf\{s:\bM\cap s\bG\neq\emptyset\}$
then $H_\bG(r)=\mbox{\rm Prob}\{R\leq r|R>0\}$ \cite{SKM95}.

For the unit ball $\bG=\bB(\vek{0},1)$ in three dimensions $H_\bB$
is called spherical contact distribution.
\index{spherical contact distribution}
The quantity $1-H_\bB(r)$ is then the distribution
function of the random distance from a randomly chosen
point in $\bP$ to its nearest neighbour in $\bM$.
The probability density
\begin{equation}
p(r)=\frac{\d}{\d r}(1-H_\bB(r))=-\frac{\d}{\d r}H_\bB(r)
\end{equation}
was discussed in \cite{sch74} as a well defined alternative
to the frequently used pore size distrubution from mercury
porosimetry.

For an oriented unit interval $\bG=\bB^1(\vek{0},1;\vek{e})$
where $\vek{e}$ is the unit vector one obtains the linear
contact distribution.
\index{linear contact distribution}
The linear contact distribution written as 
$L(r\vek{e})=\phi(1-H_{\bB^1(\vek{0},1;\vek{e})}(r))$
is sometimes called lineal path function \cite{YT98a}.
\index{lineal path function}
It is related to the chord length distribution $p_{cl}(x)$ 
defined as the probability that an interval in the intersection 
of $\bP$ with a straight line containing $\bB^1(\vek{0},1;\vek{e})$
has length smaller than $x$ \cite[p. 208]{hil95d,SKM95}.
\index{chord length distribution}

\subsection{Local Porosity Distributions}
The idea of local porosity distributions is to measure
geometric observables inside compact convex subsets
$\bK\subset\bS$, and to collect the results into empirical 
histograms \cite{hil91d}.
Let $\bK(\r,L)$ denote a cube of side length $L$ centered at the 
lattice vector $\r$.
The set $\bK(\r,L)$ is called a measurement cell.
\index{measurement cell}
A geometric observable $\f$, when measured inside a measurement
cell $\bK(\r,L)$, is denoted as $\f(\r,L)$ and called a local
observable.
\index{local observable}
An example are local Hadwiger functional densities 
$\f=\sum_{i=0}^d c_i\psi_i$ with coefficients $c_i$ 
as in Hadwigers theorem \eqref{Hadwiger}.
Here the local quermassintegrals are defined using 
\eqref{querm} as
\index{local quermassintegral}
\begin{equation}
\psi_i(\bP\cap\bK(\r,L))=
\frac{W_i(\bP\cap\bK(\r,L))}{V_d(\bK(\r,L))}
\label{lquerm}
\end{equation}
for $i=0,\ldots, d$.
In the following mainly the special case $d=3$ will be of interest. 
For $d=3$ the local porosity is defined by setting $i=0$,
\index{local porosity}
\begin{equation}
\phi(\r,L)=\psi_0(\bP\cap\bK(\r,L)) .
\label{lporosity}
\end{equation}
Local densities of surface area, mean curvature and Euler
characteristic may be defined analogously.
The local porosity distribution, defined as 
\index{local porosity distribution}
\begin{equation}
\mu(\phi;\r,L)=\gex{\de(\phi-\phi(\r,L))} ,
\label{lpd}
\end{equation}
gives the probability density to find a local porosity $\phi(\r,L)$
in the measurement cell $\bK(\r,L)$.
Here $\de(x)$ denotes the Dirac $\de$-distribution.
\index{Dirac $\de$-distribution}
The support of $\mu$ is the unit interval.
For noncubic measurement cells $\bK$ one defines
analogously $\mu(\phi;\bK)=\gex{\de(\phi-\phi(\bK))}$ where
$\phi(\bK)=\phi(\bP\cap\bK)$ is the local observable in cell~$\bK$.

The concept of local porosity distributions
\footnote{
or more generally ``local geometry distributions''
\cite{hil92a,hil95d}
}
was introduced in \cite{hil91d} and
has been generalized in two directions \cite{hil95d}.
Firstly by admitting more than one measurement cell,
and secondly by admitting more than one geometric observable.
The general $n$-cell distribution function is defined as \cite{hil95d}
\index{$n$-cell distribution}
\begin{align}
&\mu_{n;\f_1,\ldots,\f_m}
(\f_{11},\ldots,\f_{1n};\ldots;\f_{n1},\ldots,\f_{nm};\bK_1,\ldots,\bK_n)
=\nonumber\\
&\gex{\de(\f_{11}-\f_1(\bK_1))\ldots\de(\f_{1n}-\f_1(\bK_n))\ldots
\de(\f_{m1}-\f_1(\bK_1))\ldots\de(\f_{mn}-\f_m(\bK_n))}
\label{ncell}
\end{align}
for $n$ general measurement cells $\bK_1,\ldots,\bK_n$ 
and $m$ observables $\f_1,\ldots,\f_m$.
The $n$-cell distribution is the probability density to find
the values $\f_{11}$ of the local observable $\f_1$ in cell 
$\bK_1$ and $\f_{12}$ in cell $\bK_2$ and so on until $\f_{mn}$
of local observable $\f_m$ in $\bK_n$.
Definition \eqref{ncell} is a broad generalization
of \eqref{lpd}.
This generalization is not purely academic, but was 
motivated by problems of fluid flow in porous media
where not only $\psi_0$ but also $\psi_1$ becomes
important \cite{hil92a}.
Local quermassintegrals, defined in \eqref{lquerm},
and their linear combinations (Hadwiger functionals)
furnish important examples for local observables
in \eqref{ncell}, and they have recently been measured
on real sandstone samples \cite{man00}.

The general $n$-cell distribution in \eqref{ncell} is very general indeed.
It even contains $p$ from 
\eqref{stochmed} as the special case $m=1,\f_1=\phi$ and $n=N$ with
$\bK_i=\bV_i=\bK(\r_i,a)$. 
More precisely one has
\begin{equation}
\mu_{N;\phi}(\phi_1,\ldots,\phi_N;\bV_1,\ldots,\bV_N)=p(\phi_1,\ldots,\phi_N)
\end{equation}
because in that case $\phi_i=z_i=1$ if $\bV_i\in\bP$ and 
$\phi_i=z_i=0$ for $\bV\notin\bP$.
In this way it is seen that the very definition of a stochastic
geometry is related to local porosity distributions 
(or more generally local geometry distributions).
As a consequence the general $n$-cell distribution
$\mu_{n;\f_1,\ldots,\f_m}$ is again mainly of theoretical
interest, and usually unavailable for practical computations.

Expectation values with respect to $p$
have generalizations to averages with respect to $\mu$.
Averaging with respect to $\mu$ will be denoted by an overline.
In the special case $m=1,\f_1=\phi$ and $\bK_i=\bV_i=\bK(\r_i,a)$ 
with $n<N$ one finds \cite{hil95d}
\begin{align}
&\avg{\phi(\r_1,a)\cdots\phi(\r_n,a)}=\nonumber\\
&=\int_0^1\ldots\int_0^1\phi_1\cdots\phi_n
\mu_{n;\phi}(\phi_1,\ldots,\phi_n;\bV_1,\ldots,\bV_n)
\d\phi_1\cdots\d\phi_n\nonumber\\
&=\int_0^1\ldots\int_0^1\phi_1\cdots\phi_n
\mu_{N;\phi}(\phi_1,\ldots,\phi_N;\bV_1,\ldots,\bV_N)
\d\phi_1\cdots\d\phi_N\nonumber\\
&=\int_0^1\ldots\int_0^1\phi_1\cdots\phi_n
\gex{\de(\phi_1-\phi(\r_1,a))\cdots\de(\phi_N-\phi(\r_N,a))}
\d\phi_1\cdots\d\phi_N\nonumber\\
&=\gex{\phi(\r_1,a)\cdots\phi(\r_n,a)}\nonumber\\
&=\gex{\ch{\bP}(\r_1)\ldots\ch{\bP}(\r_n)}\nonumber\\
&=\M_n(\r_1,\ldots,\r_n)
\end{align}
thereby identifying the moment functions of order $n$
as averages with respect to an $n$-cell distribution.

For practical applications the $1$-cell local porosity
distributions $\mu(\r,L)$ and their analogues for other 
quermassintegrals are of greatest interest.
For a homogeneous medium the local porosity distribution obeys
\begin{equation}
\mu(\phi;\r,L)=\mu(\phi;\vek{0},L)=\mu(\phi;L)
\end{equation}
for all lattice vectors $\r$, i.e. it is independent of
the placement of the measurement cell.
A disordered medium with substitutional disorder \cite{zim82}
may be viewed as a stochastic geometry obtained by placing 
random elements at the cells or sites of a fixed regular 
substitution lattice.
\index{substitutional disorder}
For a substitutionally disordered medium
the local porosity distribution $\mu(\r,L)$ 
is a periodic function of $\r$ whose
period is the lattice constant of the substitution
lattice.
For stereological issues in the measurement of $\mu$ 
from thin sections see \cite{hil96c}.

Averages with respect to $\mu$ are denoted by an overline.
For a homogeneous medium the average local porosity is found as
\index{average local porosity}
\begin{equation}
\aphi(\r,L)=\int_0^1\mu(\phi;\r,L)\d\phi=\gex{\phi}=\aphi
\label{alporosity}
\end{equation}
independent of $\r$ and $L$.
The variance of local porosities for a homogeneous medium 
defined in the first equality
\index{variance of local porosities $\si^2$}
\begin{align}
\si^2(L)=\avg{(\phi(L)-\aphi)^2}&=
\int_0^1(\phi(L)-\aphi)^2\mu(\phi;L)\d\phi\nonumber\\
&=\frac{1}{L^3}\gex{\phi}(1-\gex{\phi})\left(
1+\frac{2}{L^3}
\sum_{{\r_i,\r_j\in\bK(\r_0,L)}\atop{i\neq j}}
G(\r_i-\r_j)
\right)
\label{lpvariance}
\end{align}
is related to the correlation function as given in the
second equality \cite{hil95d}.
The skewness of the local porosity distribution is defined
as the average
\index{skewness of local porosities}
\begin{equation}
\ka_3(L) = \frac{\avg{(\phi(L)-\aphi)^3}}{\si(L)^3}
\label{lpskewness}
\end{equation}

The limits $L\to 0$ and $L\toi$ of small and large
measurement cells are of special interest.
In the first case one reaches the limiting resolution 
at $L=a$ and finds for a homogeneous medium \cite{hil91d,hil95d}
\begin{equation}
\mu(\phi;a)=\aphi\de(\phi-1)-(1-\aphi)\de(\phi) .
\label{lim1}
\end{equation}
The limit $L\toi$ is more intricate because it requires
also the limit $\bS\to\bR^3$.
For a homogeneous medium \eqref{lpvariance}
shows $\si(L)\to 0$ for $L\to 0$ and this suggests
\begin{equation}
\mu(\phi,L\toi)=\de(\phi-\aphi).
\label{lim2}
\end{equation}
For macroscopically heterogeneous media, however,
the limiting distribution may deviate from this result
\cite{hil95d}.
If \eqref{lim2} holds then in both limits the geometrical
information contained in $\mu$ reduces to the single number
$\aphi=\gex{\phi}$.
If \eqref{lim1} and \eqref{lim2} hold there exists a special
length scale $L^*$ defined as
\begin{equation}
L^*=\min\{L:\mu(0;L)=\mu(1;L)=0\}
\end{equation}
at which the $\de$-components at $\phi=0$ and $\phi=1$ 
vanish.
The length $L^*$ is a measure for
the size of pores.

The ensemble picture underlying the definition of a stochastic
medium is an idealization.
In practice one is given only a single realization and has
to resort to an ergodic hypothesis for obtaining an estimate
of the local porosity distributions.
The local porosity distribution may then be estimated by
\index{local porosity distribution}
\begin{equation}
\muw(\phi;L)=\frac{1}{m}\sum_{\r}\de(\phi-\phi(\r,L))
\end{equation}
where $m$ is the number of placements of the measurement
cell $\bK(\r,L)$.
Ideally the measurement cells should be far apart or at
least nonoverlapping, but in practice this restriction
cannot be observed because the samples are not large enough.
The use of $\muw$ instead of $\mu$ can lead to deviations due to
violations of the ergodic hypothesis or simply due to oversampling
the central regions of $\bS$ \cite{hil98a,hil99c}.

\subsection{Local Percolation Probabilities}
Transport and propagation in porous media are controlled
by the connectivity of the pore space.
Local percolation probabilities characterize the connectivity
\cite{hil91d}.
Their calculation requires a threedimensional pore space 
representation, and early results were restricted to samples
reconstructed laboriously from sequential thin sectioning \cite{hil96f}
In this section a relationship between the Euler characteristic
and the local percolation probabililties is established for the
first time.

Consider the functional $\La:\cK\times\cK\times\cR\to\bZ_2=\{0,1\}$
defined by
\begin{equation}
\La(\KA,\KE;\bP\cap\bS)=
\begin{cases}
1 : \text{~if~} \KA\leadsto\KE \text{~in~} \bP\\
0 : \text{~otherwise}
\end{cases}
\end{equation}
where $\KA\subset\bR^3,\KE\subset\bR^3$ are two compact 
convex sets with $\KA\cap(\bP\cap\bS)\neq\emptyset$ and 
$\KE\cap(\bP\cap\bS)\neq\emptyset$,
and ``$\KA\leadsto\KE$ in $\bP$'' means that 
there is a path connecting $\KA$ and $\KE$ that lies 
completely in $\bP$.
In the examples below the sets $\KA$ and $\KE$ correspond to 
opposite faces of the sample, but in general other choices
are allowed.
Analogous to $\La$, which is defined for the whole sample, one defines 
for a measurement cell
\begin{equation}
\La_\al(\r,L)=\La(\KA_\al,\KE_\al;\bP\cap\bK(\r,L))=
\begin{cases}
1 : \text{~if~} \KA_\al\leadsto\KE_\al \text{~in~} \bP\\
0 : \text{~otherwise}
\end{cases}
\end{equation}
where $\al=x,y,z$ and $\KA_x,\KE_x$ denote those two faces of 
$\bK(\r,L)$ that are normal to the $x$ direction.
Similarly $\KA_y,\KE_y,\KA_z\KE_z$ denote the faces of $\bK(\r,L)$
normal to the $y$- and $z$-directions.
Two additional percolation observables $\La_3$ and $\La_c$ 
are introduced by
\begin{align}
\La_3(\r,L)&=\La_x(\r,L)\La_y(\r,L)\La_z(\r,L)\\
\La_c(\r,L)&=\sgn(\La_x(\r,L)+\La_y(\r,L)+\La_z(\r,L)) .
\end{align}
$\La_3$ indicates that the cell is percolating in
all three directions while $\La_c$ indicates
percolation in $x$- or $y$- or $z$-direction.
The local percolation probabilities are defined as
\index{local percolation probability}
\begin{equation}
\la_\al(\phi;L)=
\frac{\sum_\r\La_\al(\r,L)\de_{\phi,\phi(\r,L)}}
{\sum_\r\de_{\phi,\phi(\r,L)}}
\end{equation}
where
\begin{equation}
\de_{\phi,\phi(\r,L)}=\begin{cases}
1 : \text{~if~}\phi=\phi(\r,L)\\
0 : \text{~otherwise}.
                     \end{cases}
\end{equation}
The local percolation probability $\la_\al(\phi;L)$ gives
the fraction of measurement cells of sidelength $L$ with
local porosity $\phi$ that are percolating in the ``$\al$''-direction.
The total fraction of cells percolating along the ``$\al$''-direction
is then obtained by integration
\begin{equation}
p_\al(L)=\int_0^1\mu(\phi;L)\la_\al(\phi;L)\d\phi .
\label{pL1}
\end{equation}
This geometric observable is a quantitative measure for the number
of elements that have to be percolating if the pore
space geometry is approximated by a substitutionally 
disordered lattice or network model.
Note that neither $\La$ nor $\La_\al$ are additive
functionals, and hence local percolation probabilities
are not covered by Hadwigers theorem.

It is interesting  that there is a relation between
the local percolation probabilities and the local Euler
characteristic $V_0(\bP\cap\bK(\r,l))$.
\index{local Euler characteristic}
The relation arises from the observation that the voxels 
$\bV_i$ are closed, convex sets, and hence for any two voxels 
$\bV_i,\bV_j$ the Euler characteristic of their intersection
\begin{equation}
V_0(\bV_i\cap\bV_j)=\begin{cases}
1 : \text{~if~}\bV_i\cap\bV_j\neq\emptyset\\
0 : \text{~if~}\bV_i\cap\bV_j=\emptyset
                   \end{cases}
\end{equation}
indicates whether two voxels are nearest neighbours.
$\!$A measurement cell $\bK(\r,L)$ contains $L^3$ voxels.
It is then possible to construct a $(L^3+2)\times(L^3+2)^2$-matrix
$B$ with matrix elements
\begin{align}
(B)_{i\;(i,j)}&=V_0(\bV_i\cap\bV_j)\\
(B)_{i\;(j,i)}&=-V_0(\bV_i\cap\bV_j)
\end{align}
where $i,j\in\{0,1,\ldots,L^3,\infty\}$ and the
sets $\bV_0=\KA$ and $\bV_\infty=\KE$ are two opposite
faces of the measurement cell.
The rows in the matrix $B$ correspond to voxels while
the columns correspond to voxel pairs.
Define the matrix $A=BB^T$
where $B^T$ is the transpose of $B$.
The diagonal elements $(A)_{ii}$ give the number of 
voxels to which the voxel $\bV_i$ is connected.
A matrix element $(A)_{ij}$ differs from zero
if and only if $\bV_i$ and $\bV_j$ are connected.
Hence the matrix $A$ reflects the local
connectedness of the pore space around a single voxel.
Sufficiently high powers of $A$ provide information about the global 
connectedness of $\bP$.
One finds
\begin{equation}
\La(\KA,\KE;\bP\cap\bK(\r,L))=
\sgn\left(|(A^m)_{0\infty}|\right)
\end{equation}
where $(A^m)_{0\infty}$ is the matrix element in the upper right
hand corner and $m$ is arbitrary subject to the condition $m>L^3$.
The set $\bP\cap\bK(\r,L)$ can always be decomposed uniquely
into pairwise disjoint connectedness components (clusters)
$\bB_i$ whose number is given by the rank of $B$.
Hence
\begin{equation}
V_0(\bP\cap\bK(\r,L))=\sum_{i=1}^{\mbox{\scriptsize rank}B}V_0(\bB_i)
\end{equation}
provides an indirect connection between the local 
Euler characteristic and the local percolation probabilities
mediated by the matrix $B$.
\footnote{For percolation systems it has been conjectured that
the zero of the Euler characteristic as a function of
the occupation probability is an approximation
to the percolation threshold  \cite{MW91}})

The theoretical concepts for the geometric characterization
of porous media discussed here are also useful in effective
medium calculations of transport parameters such as conductivity
or permeability \cite{hil91d,hil92a,hil95d}.
The resulting parameterfree predictions agree well with
the exact result \cite{hil98h,hil99a}.

\section{Application Example}
The purpose of this section is to show that the previous
theoretical framework can be used directly for predicting
transport in porous media quantitatively and without
free macroscopic fit parameters.
The theory presented above allows a quantitative micro-macro
transition in porous media and opens the possibility
to determine the elusive ``representative elementary
volume'' needed for macroscopic theories in a
quantitative and property specific manner.

\begin{figure}
\psfrag{pl3}{\Large $p_3(L)$}
\epsfig{figure=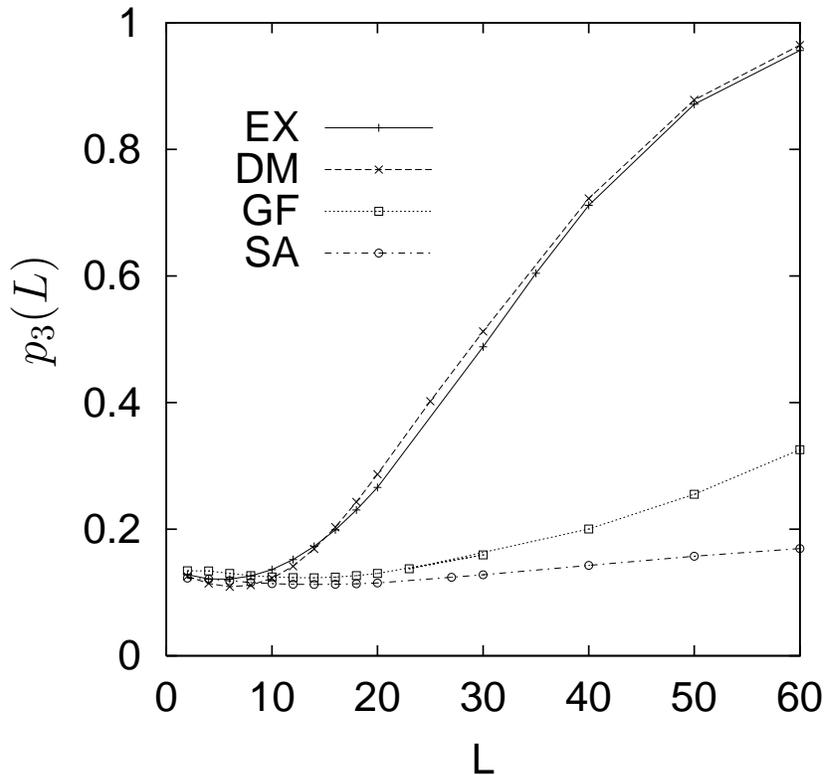,angle=0,width=16cm}
\caption{Total fraction of percolating cells for Fontainebleau
sandstone (\EX) and three of its models (\DM,\GF,\SA) as
discussed in Ref. \cite{hil99c}} 
\label{fig}
\end{figure}

In Reference \cite{hil99c} a fully threedimensional experimental 
sample of Fontainebleau sandstone was compared with three
geometric models, some of which had not only the same porosity
and specific internal surface area but also the same correlation
function $G(r)$.
A large number of the geometric quantities discussed above
was calculated in \cite{hil99c}.
For a discussion on the influence contact of distrtributions see
\cite{hil00g}.
The total fraction of percolating cells, defined in eq.
\eqref{pL1} above, was measured in
\cite{hil99c} for Fontainebleau and some of its models.
Figure \ref{fig} shows the total fraction of percolating
cells as a function of length scale (side length of 
measurement cells).

It turns out that the quantity $p_3(L)$ displayed in
Figure \ref{fig} correlates very well with transport
properties such as the hydraulic permeability.
Recently transport properties such as the 
permesbilities and formation factors of the Fontainebleau 
sandstone and its geometries were calculated
numerically exactly by solving the appropriate
microscopic equations of motion on the computer
\cite{man00}.
Some of the results are summarized in Table \ref{tab:trans} below.

\begin{table}
  \caption{Physical transport properties of Fontainebleau sandstone
and three geometric models for it (see \cite{hil99c}).
$\sigma_{ii}$ is the conductivity in the direction $i=x,y,z$
in units of $10^{-3}\sigma_\bP$, where $\sigma_\bP$ is the
conductivity of thematerial filling the pore space.
$k_{ii}$ is the permeability in the direction $i=x,y,z$ in mD.}
\label{tab:trans}
  \begin{tabular}{lllll}
       & \EX & \DM & \SA & \GF\\
    \hline
$k_{zz}$[mD]   & 692 & 923 & 35 & 34\\
$k_{yy}$[mD]   & 911 & 581 & 22 & 35\\
$k_{xx}$[mD]   & 790 & 623 & 20 & 36\\
$\sigma_{zz}$[$10^{-3}\sigma_\bP$]    & 18.5 & 26.2  & 1.35 & 2.05 \\
$\sigma_{yy}$[$10^{-3}\sigma_\bP$]    & 21.9 & 17.0  & 0.87 & 1.97 \\
$\sigma_{xx}$[$10^{-3}\sigma_\bP$]    & 20.5 & 17.1  & 0.96 & 1.98 
  \end{tabular}
\end{table}

One sees from Table \ref{tab:trans} that while \EX~ and \DM~ 
are very similar in their permeabilities and formation
factors the samples \EX~ and \GF~ have significantly
lower values with \GF~ being somewhat higher than \SA.
The same relationship is observed in Figure \ref{fig}
for the percolation properties.
These results show that the purely geometrical local 
percolation probabilities correlate surprisingly well 
with hydraulic permeability and electrical conductivity
that determine physical transport.

\vspace*{2cm}
ACKNOWLEDGEMENT:
Many of the results discussed in this paper were obtained in earlier
cooperations with B. Biswal, C. Manwart, J. Widjajakusuma, 
P.E. {\O}ren, S. Bakke, and J. Ohser. 
I am grateful to all of them, and to the Deutsche 
Forschungsgemeinschaft as well as Statoil A/S Norge
for financial support.

\newpage

\end{document}